# Some Preliminary Result About the Inset Edge and Average Distance of Trees

**M. H. Khalifeh*, A.-H. Esfahanian**

Department of Computer Science and Engineering Michigan State University East Lansing, MI 48824, USA

**Abstract:** An added edge to a graph is called an *inset edge*. Predicting $k$ inset edges which minimize the average distance of a graph is known to be NP-Hard. However, when $k = 1$ the complexity of the problem is polynomial. In this paper, some tools for a precise analysis of the problem for the trees are established. Using the tools, we can avoid using the distance matrix. This leads to more efficient algorithms and a better analysis of the problem. Several applications of the tools as well as a tight bound for the change of average distance when an inset edge is added to a tree are presented.

## Introduction:

Average distance, degree distribution and clustering coefficient are the three most robust measures of network topology. The average distance of a graph is the average of distances between every pair of vertices with finite distances [1]. The sensitivity of average distance of a network, both after missing a link [5] or adding a link is interesting for the researchers [3]-[4]. The efficiency of mass transfer in a metabolic network can be judged by studying its average path length [5]. On the other hand, finding $k$ new links which minimize the average distance is an NP-Hard problem [7]. According to [7] there exists a polynomial time algorithm for finding an edge with maximum change on the average distance.

*Corresponding author's Email: khalife8@msu.edu

In this paper we base some tools which helps to analyze the change of average distance of a tree after adding one edge. After adding one edge to a tree some distances change. The tools let us to avoid re-calculating the distances with no changes. A tree $T$ on $n$ vertices has $\frac{n^2-3n}{2} = |E(T^c)|$ inset edges. Using the tools, we can limit, the search space of inset edge(s), with maximum change on the average distance. Other applications of the tools such as finding bounds for the average distance change are achieved.

From here on we switch to the relevant standard graph theory terminology to state notations, definitions and results.

## Definitions, Notations and Results:

For a graph $G$ and $u, v \in V(G)$, $d_G(u,v)$ denotes the distance between $u$ and $v$ which is the length of a shortest path between $u$ and $v$, if there is a path between them, and is equal to infinity otherwise. Using this, for $A, B \subseteq V(G)$ let:

$$d_G(A,B) = \frac{1}{2} \sum_{(a,b) \in A \times B} d_G(a,b) \tag{1}$$

Thus, we have the sum of distances between all pair of vertices as,

$$D(G) = d_G\big(V(G), V(G)\big).$$

$D(G)$ is also known as Wiener number of $G$ [10]. Therefore, the average distance of a graph is equal to $AD(G) = D(G) \cdot R^{-1}$ where $R$ denotes the number of pairs of vertices of $G$ with finite distances. When $G$ is a connected simple graph, $R = \binom{|V(G)|}{2}$.

For ease we may use $D(G)$ instead of $AD(G)$ since the difference is a constant coefficient in our case.

For a given graph, $G \neq K_n$ (complete graph on $n$ vertices) and $Z \subseteq E(\overline{G})$, let ${}_ZG = G + Z$. Therefore, for a tree $T$ and $xy \in E(\overline{T})$ , ${}_{xy}T$ is a unicyclic graph. In this case we call the edge $xy$ the *inset edge*. When we know the length of the cycle of ${}_{xy}T$ is $k$ we indicate it by ${}_{xy}T^k$.

**Definition 1:** For ${}_ZG$ we define the $D'$and $AD'$indices as follows:

$$\boldsymbol{D'({}_ZG) = \frac{D(G) - D({}_ZG)}{|Z|}},$$

and

$$\boldsymbol{AD'({}_ZG) = \frac{AD(G) - AD({}_ZG)}{|Z|}}.$$

Note that in ${}_{xy}T$, $|Z| = |\{xy\}| = 1$. Therefore $D'(T) = D(T) - D({}_{xy}T)$. Moreover, for a connected graph $G$, $\binom{|V(G)|}{2}AD'(G) = D'(G)$. Informally the prime notation denotes change of average distances after a change on the graph.

Suppose $C$ is the cycle of ${}_{xy}T^k$, we define,

$$C_x = \{\, v \in V(C) \mid d_T(x,v) < d_T(y,v) \,\},$$

$$C_y = \{\, v \in V(C) \mid d_T(x,v) > d_T(y,v) \,\},$$

$$C_M = \{\, v \in V(C) \mid d_T(x,v) = d_T(y,v) \,\}.$$

Note that $|C_x| = |C_y| = \left\lfloor \frac{k}{2} \right\rfloor$. $|C_M| = 0$ when $k$ is even and $|C_M| = 1$ otherwise. Now we propose an indexing of $C_x$ to $x_i$'s, $1 \leq i \leq \left\lfloor \frac{k}{2} \right\rfloor$, such that $d_T(x, x_i) = i - 1$. Similarly, we index the elements of $C_y$ to $y_i$'s. Moreover, if $C_M$ has its single element we call it $x_M$. As we observe the indexing is unique and partitions the vertices of $C$, the cycle of ${}_{xy}T$. On the other hand, for $v \in V(C)$ suppose $T_v$ to be the maximal subtree of ${}_{xy}T$ such that,

$$V(T_v) \cap V(C) = \{v\}.$$

Informally $T_v$ is the sub-tree attached to the vertex $v \in V(C)$ of the cycle of ${}_{xy}T$. We denote the number of vertices of $T_v$, $|V(T_v)|$, by $w_v$.

Using the notations and definitions we have the following theorem which has a critical role for our analysis and reduction of time complexity of the actual calculations.

**Theorem 1: For a tree $\boldsymbol{T}$,**

$$\boldsymbol{D'\left({}_{xy}T^k\right) = \sum_{\substack{(u,v)\in C_x\times C_y \\ d_T(u,v)>\frac{k}{2}}} (2d_T(u,v) - k).w_u.w_v}$$

**Proof:** Suppose $C$ is the cycle of ${}_{xy}T^k$. Using the definition of $D$ and our notations,

$$\begin{aligned} D\left({}_{xy}T^k\right) = & \sum_{\{u,v\}\subseteq V(C_x\,\cup\, C_M)} d_{{}_{xy}T^k}(V(T_u),V(T_v)) + \sum_{\{u,v\}\subseteq V\left(C_y\,\cup\, C_M\right)} d_{{}_{xy}T^k}(V(T_u),V(T_v)) \\ & + \sum_{v\in V(C)} d_{{}_{xy}T^k}(V(T_v),V(T_v)) + \sum_{\substack{(u,v)\in C_x\times C_y \\ d_T(u,v)\le\frac{k}{2}}} d_{{}_{xy}T^k}(V(T_u),V(T_v)) \\ & + \sum_{\substack{(u,v)\in C_x\times C_y \\ d_T(u,v)>\frac{k}{2}}} d_{{}_{xy}T^k}(V(T_u),V(T_v)) \end{aligned} \tag{2}$$

Looking at the updated distances from $T$ to ${}_{xy}T^k$ one can check that,

$$d_{{}_{xy}T^k}(V(T_u),V(T_v)) = d_T(V(T_u),V(T_v)) \qquad \text{except for} \qquad \begin{matrix} (u,v) \in C_x \times C_y \\ d_T(u,v) > \frac{k}{2} \end{matrix}.$$

Therefore,

$$D(T) - D({}_{xy}T^k) = \sum_{\substack{(u,v)\in C_x\times C_y \\ d_T(u,v)>\frac{k}{2}}} d_T(V(T_u),V(T_v)) - \sum_{\substack{(u,v)\in C_x\times C_y \\ d_T(u,v)>\frac{k}{2}}} d_{{}_{xy}T^k}(V(T_u),V(T_v)). \quad (3)$$

In addition, for every $u \in V(T_{x_i})$, $v \in V(T_{y_i})$, if $d_T(x_i, y_j) > \frac{k}{2}$,

$$d_T(x_i, y_j) > d_{{}_{xy}T^k}(x_i, y_j). \quad (4)$$

$$d_T(u, x_i) = d_{{}_{xy}T^k}(u, x_i), \quad (5)$$

$$d_T(v, y_i) = d_{{}_{xy}T^k}(v, y_i), \quad (6)$$

More precisely for the given condition we can use the new paths formed after adding the inset edge, $xy$, which gives us,

$$d_T(x_i, y_i) - d_{{}_{xy}T^k}(x_i, y_i) = 2.d_T(x_i, y_i) - k .$$

Since we have $|T_{x_i}| = w_{x_i}$ and $|T_{y_i}| = w_{y_i}$ so for $d(x_i, y_i) > \frac{k}{2}$ using (4) to (6),

$$d_T\left(V(T_{x_i}), V(T_{y_i})\right) - d_{{}_{xy}T^k}\left(V(T_{x_i}), V(T_{y_i})\right) = (2.d_T(x_i, y_i) - k).w_{x_i}.w_{y_i}. \quad (7)$$

The equations (3) and (7) leads us to,

$$D'({}_{xy}T^k) = \sum_{\substack{(u,v)\in C_x\times C_y \\ d_T(u,v)>\frac{k}{2}}} (2d_T(u,v) - k).w_u.w_v.$$

This completes the proof. ■

As a useful example of last theorem, $D'({}_{uv}T^3) = w_u.w_v$.

**Corollary 1:** For a tree $T$,

$$\boldsymbol{max_{xy\in E(T^c)} D'({}_{xy}T) = min_{xy\in E(T^c)} D({}_{xy}T^k)}.$$

**Proof:** By the definition 1 and the fact that $D(T)$ is a constant here. ■

**Remark 1:** As defined, when we introduce a graph, ${}_{xy}T^k$ we mean that the inset edge is $xy$ which makes a $k$-cycle for ${}_{xy}T^k$. Since in the indexing of the cycle of ${}_{xy}T^k$, $x = x_1$ and $y = y_1$ so ${}_{xy}T^k = {}_{x_1y_1}T^k$. Moreover, we abuse the indexing of the cycle of ${}_{xy}T^k$ to say that the inset edge of ${}_{x_2y_1}T^{k-1}$ is $x_2y_1$. Also, ${}_{x_0y_1}T^{k+1}$ means its inset edge is $x_0y_1$ where $x_0$ is a neighbor of $x_1$ except from $x_2$. A similar argument applies for ${}_{x_iy_j}T^{k-i-j+2}$ regarding the indexing of the cycle of ${}_{xy}T^k$. ▲

To use the benefits of Theorem 1 we require to create some tools to be able to use some matrix calculations. For ease hereafter let $k' = \left\lfloor \frac{k}{2} \right\rfloor$.

Suppose we are given a ${}_{xy}T^k$. We associate the vectors ${}^{x}_{xy}\boldsymbol{W}$ or ${}^{x}_{xy}\boldsymbol{W}^k = [{}^{x}_{xy}w_i]$ to it which is a $k'$−vector and ${}^{x}_{xy}w_i = w_{x_i} = |T_{x_i}|$. Similarly, we define ${}^{y}_{xy}\boldsymbol{W}$ or ${}^{y}_{xy}\boldsymbol{W}^k = [{}^{y}_{xy}w_i]$ where ${}^{y}_{xy}w_i = w_{y_i} = |T_{y_i}|$. Finally, we associate a matrix ${}_{xy}\boldsymbol{W}$ to ${}_{xy}T$ as follows,

$$ {}_{xy}\boldsymbol{W} = {}^{x}_{xy}\boldsymbol{W} \times ({}^{y}_{xy}\boldsymbol{W})^t. $$

Therefore, in a ${}_{xy}\boldsymbol{W} = [w_{ij}]$, $w_{ij} = {}^{x}_{xy}w_i \cdot {}^{y}_{xy}w_j = w_{x_i}.w_{y_j}$. Next, we introduce the matrix $\boldsymbol{F}_k$. The matrix $\boldsymbol{F}_k$ is a $k' \times k'$ matrix as follows,

$$ \boldsymbol{F}_k = \begin{cases} \boldsymbol{D}_k + \boldsymbol{O}_k & k \text{ is odd,} \\ \boldsymbol{D}_k & \text{otherwise.} \end{cases} $$

where $\boldsymbol{D}_k = [d_{ij}]$ and $\boldsymbol{O}_k = [o_{ij}]$ are also $k' \times k'$ matrices as follows,

$$ d_{ij} = \begin{cases} 2(k' - i - j + 1) & i + j \leq k', \\ 0 & \text{otherwise.} \end{cases} $$

and

$$o_{ij} = \begin{cases} 1 & i + j - 1 \le k', \\ 0 & \text{otherwise.} \end{cases}$$

For more resolution,

$$\boldsymbol{F}_k = \begin{bmatrix} 2k'-1 & 2k'-3 & 2k'-5 & & \dots & 1 \\ 2k'-3 & 2k'-5 & & \dots & 1 & 0 \\ 2k'-5 & & \dots & 1 & 0 & 0 \\ & & \vdots & & & \\ & \vdots & 1 & & & \vdots \\ \vdots & 1 & 0 & & \dots & 0 \\ 1 & 0 & 0 & \dots & 0 & 0 \end{bmatrix} \quad k \text{ is odd}$$

$$\boldsymbol{F}_k = \begin{bmatrix} 2k'-2 & 2k'-4 & 2k'-6 & & & \dots & 2 & 0 \\ 2k'-4 & 2k'-6 & & & \dots & 2 & 0 & 0 \\ 2k'-6 & \vdots & & \dots & 2 & 0 & 0 & 0 \\ & & \vdots & & & & & 0 \\ & \vdots & 2 & & & & & \\ \vdots & 2 & 0 & & & & & \vdots \\ 2 & 0 & 0 & & & & \dots & 0 \\ 0 & 0 & 0 & & & \dots & 0 & 0 \end{bmatrix} \quad k \text{ is even}$$

$$\boldsymbol{O}_k = \begin{bmatrix} 1 & 1 & 1 & & & \dots & 1 \\ 1 & 1 & & & \dots & 1 & 0 \\ 1 & & & \dots & 1 & 0 & 0 \\ & & & \vdots & & & \\ & & \vdots & 1 & & & \vdots \\ \vdots & & 1 & 0 & & \dots & 0 \\ 1 & & 0 & 0 & \dots & 0 & 0 \end{bmatrix}$$

$${}_{xy}\boldsymbol{W} = \begin{bmatrix} w_{x_1} \\ w_{x_2} \\ \vdots \\ w_{x_{k'}} \end{bmatrix} \begin{bmatrix} w_{y_1} & w_{y_2} & \dots & w_{y_{k'}} \end{bmatrix} = \begin{bmatrix} w_{x_1}.w_{y_1} & w_{x_1}.w_{y_2} & w_{x_1}.w_{y_3} & \dots & w_{x_1}.w_{y_{k'}} \\ w_{x_2}.w_{y_1} & w_{x_2}.w_{y_2} & \vdots & \dots & w_{x_2}.w_{y_{k'}} \\ w_{x_3}.w_{y_1} & \vdots & & \dots & w_{x_3}.w_{y_{k'}} \\ \vdots & & \dots & & \vdots \\ w_{x_{k'-2}}.w_{y_1} & \dots & & \vdots & w_{x_{k'-2}}.w_{y_{k'}} \\ w_{x_{k'-1}}.w_{y_1} & \dots & \vdots & w_{x_{k'-1}}.w_{x_{k'-1}} & w_{x_{k'-1}}.w_{y_{k'}} \\ w_{x_{k'}}.w_{y_1} & \dots & w_{x_{k'}}.w_{y_{k'-2}} & w_{x_{k'}}.w_{y_{k'-1}} & w_{x_{k'}}.w_{y_{k'}} \end{bmatrix}$$

In fact, $\boldsymbol{F}_k$ is an upper-triangular matrix which provides the coefficients of $w_u.w_v$ in the formula of $D'$ in the Theorem 1.

To express our next result, we remind the Hadamard multiplication and norm one of matrices as follows.

If $\boldsymbol{A} = [a_{ij}]$ is a matrix then the norm one of $A$ is the following:

$$\|\boldsymbol{A}\| = \sum_{i,j} |a_{ij}| \,.$$

The Hadamard product of two matrices $\boldsymbol{A} = [a_{ij}]$ and $\boldsymbol{B} = [b_{ij}]$ with the same dimensions is an element-wise product which is defined as follows,

$$\boldsymbol{A} \odot \boldsymbol{B} = [a_{ij}.b_{ij}].$$

Note also that $e_i$ denotes the $i$th standard vector of proper dimension on the text.

**Lemma 1:** For a tree $T$ we have:

$$\boldsymbol{D}\left({}_{xy}\boldsymbol{T}^{\boldsymbol{k}}\right) = \left\|\boldsymbol{F}_{\boldsymbol{k}} \odot {}_{xy}\boldsymbol{W}\right\|.$$

**Proof:** Suppose $C$ is the cycle of ${}_{xy}T^k$ and $u, v \in V(C)$. According to the definition, the $ij$'th entry of ${}_{xy}W$ is equal to $w_u.w_v$. And the $ij$'th entry of $F_k$ for $d_T(u,v) > \frac{k}{2}$ is equal to $(2d_T(u,v) - k)$ and is zero otherwise. Therefore, using the definition of norm one and Hadamard multiplication:

$$\left\|\boldsymbol{F}_k \odot {}_{xy}\boldsymbol{W}\right\| = \sum_{\substack{(u,v)\in C_x\times C_y \\ d_T(u,v)>\frac{k}{2}}} (2d_T(u,v) - k).w_u.w_v$$

By Theorem 1 the RHS of the above equation is equal to $D\left({}_{xy}T^k\right)$. ■

The next theorem gives the lower and upper bound of $D\left({}_{xy}T^k\right)$ over the set of all trees with $n$ vertices. The upper bound is interesting since for a tree $T$ with $n$ vertices $(n-1)^2 \leq D(T) \leq \binom{n}{3}$, see [9]. We underline that $P_n$ and $S_n$ denotes a path and a star over $n$ vertices respectively. Moreover $e_i$ is the $i$'th unit standard vector.

**Theorem 2:** For a tree $T$ with $n$ vertices,

$$1 \leq W'({}_{xy}T) \leq \frac{n^3}{16} - \frac{n^2}{32} - \frac{9n}{8} + 2.$$

Moreover, the lower bound holds if and only if $T$ has two leaves with the distance equal to 2. The upper bound holds if and only if the following holds,

1) $n \equiv 0 \bmod 8, n \geq 16,$
2) $T$ is formed by attaching an arbitrary vertex of an arbitrary tree with $\frac{n}{4}$ vertices to a leaf of a $P_{\frac{n}{2}+3}$ and also attaching an arbitrary vertex of another arbitrary tree with $\frac{n}{4}$ vertices to the other leaf of $P_{\frac{n}{2}+3}$.
3) $xy$ corresponds to the leaves of $P_{\frac{n}{2}+3}$ in the tree formed in the previous step.

**Proof :** To prove the upper bound, suppose we are given the graph ${}_{xy}T^k$ with some non-trivial $T_{x_i}$, where $x_i$ belongs to $C_x/\{x_1\}$ and $|V(T)| = n$. Accordingly, $i \neq 1$. Moreover let $C$ to be the cycle of ${}_{xy}T^k$. Detach the elements of $V(T_{x_i})/\{x_i\}$ and attach arbitrarily to the vertices of $T_{x_1}$ and call the new graph, $G$. By lemma 1 we can see that if $\left|V(T_{x_i})/\{x_i\}\right| = w$,

$$\begin{aligned} D'(G) &= \left\| \boldsymbol{F}_k \circledcirc \left( {}^{x}_{xy}\boldsymbol{W} + w.e_1 - w.e_i \right). \left( {}^{y}_{xy}\boldsymbol{W} \right)^t \right\| \\ &= D'\left({}_{xy}T^k\right) + \left\| \boldsymbol{F}_k \circledcirc (w.e_1). ({}^{y}_{xy}\boldsymbol{W})^t \right\| - \left\| \boldsymbol{F}_k \circledcirc (w.e_i). ({}^{y}_{xy}\boldsymbol{W})^t \right\|. \quad (8) \end{aligned}$$

This is straightforward to check that,

$$\left\| \boldsymbol{F}_k \circledcirc (w.e_1). ({}^{y}_{xy}\boldsymbol{W})^t \right\| > \left\| \boldsymbol{F}_k \circledcirc (w.e_i). ({}^{y}_{xy}\boldsymbol{W})^t \right\|.$$

Therefore by (8),

$$D'(G) > D'\left({}_{xy}T^k\right).$$

Consequently, since $i$ was arbitrary and $i \neq 1$:

1. $D'({}_{xy}T^k)$ is maximum $\Leftrightarrow$ $\forall\ T_{x_i}$, $2 \le i \le k'$, is trivial.

2. Like wisely, $D({}_{xy}T^k)$ is maximum $\Leftrightarrow$ $\forall\ T_{y_i}$, $2 \le i \le k'$, is trivial.

3. Since the matrix ${}_{xy}\boldsymbol{W}$ is independent of $w_{x_M}$,

   $D({}_{xy}T^k)$ is maximum $\Leftrightarrow$ $T_{x_M}$ is trivial.

4. Using the facts 1 To 3,

   I: $D({}_{xy}T^k)$ is maximum $\Leftrightarrow \forall v \in V(C)/\{x,y\}, |T_v| = w_v = 1$.

   II: Since $|C| = k$ and $|V(T)| = n$ so $D({}_{xy}T^k)$ is maximum $\Leftrightarrow$

$$w_x + w_y = n - k + 2. \tag{9}$$

5. Using the lemma 1 and (9) we can see that $D({}_{xy}T^k)$ is maximal $\Leftrightarrow$

$$w_x.w_y = \left\lfloor \frac{n-k+2}{2} \right\rfloor \left\lceil \frac{n-k+2}{2} \right\rceil (k-2).$$

Consequently, by the facts 1 to 5 and the Lemma 1, if $D({}_{xy}T^k)$ is maximal and $k \equiv 2 \bmod 4$:

$$D({}_{xy}T^k) = \left\lfloor \frac{n-k+2}{2} \right\rfloor \left\lceil \frac{n-k+2}{2} \right\rceil (k-2) + \frac{1}{4}(k-2)(k-4)(n-k+2) + \frac{1}{2}(k-4)(k-6) - \frac{1}{8}(k-6)(k-2)\,, \tag{10}$$

and for $k \equiv 0 \bmod 4$:

$$D({}_{xy}T^k) = \left\lfloor \frac{n-k+2}{2} \right\rfloor \left\lceil \frac{n-k+2}{2} \right\rceil (k-2) + \frac{1}{4}(k-2)(k-4)(n-k+2) + \frac{1}{2}(k-4)(k-6) - \frac{1}{8}(k-4)^2\,, \tag{11}$$

and for odd $k$'s:

$$D'({}_{xy}T^k) = \left\lfloor\frac{n-k+2}{2}\right\rfloor\left\lceil\frac{n-k+2}{2}\right\rceil(k-2) + \frac{1}{4}(k-3)^2(n-k+2) + \frac{1}{2}(k-5)^2 - \frac{1}{8}(k-5)(k-3). \quad (12)$$

Since $n = |V(T)|$ is a constant so $D'({}_{xy}T^k)$ in (10) to (12) is a function of $k = |C|$. To find the maximum of $D'({}_{xy}T^k)$ we can take derivative of $D'({}_{xy}T^k)$ respect to $k$ for each case of (11) to (12) and find the critical points. To do this we require to consider 2 cases. First if $\frac{n-k+2}{2}$ is an integer then the roots of $\frac{\partial D'({}_{xy}T^k)}{\partial k}$ in the (10) to (12) are equal to the following respectively,

$$k_1 = \frac{n^2+2n-24}{2n-7}, \qquad k_2 = \frac{n^2-2n-24}{2n-7} \quad \text{and} \quad k_3 = \frac{n^2-2n-25}{2n-7}.$$

Second if $\frac{n-k+2}{2}$ is not an integer then the critical points of $\frac{\partial D'({}_{xy}T^k)}{\partial k}$ in the (10)-(12) are respectively,

$$k_1 = \frac{n^2+2n-26}{2n-7}, \qquad k_2 = \frac{n^2-2n-25}{2n-7} \quad \text{and} \quad k_3 = \frac{n^2-2n-26}{2n-7}.$$

Now for $k_i$ , $i = 1 \text{ or } 2 \text{ or } 3$ we require to use the values $\lfloor k_i \rfloor \text{ or } \lceil k_i \rceil$ to find the maximum of $D'({}_{xy}T^k)$ in their respective equations. After doing so, we observe that the maximum reachable value is equal to

$$\frac{n^3}{16} - \frac{n^2}{32} - \frac{9n}{8} + 2.$$

This proves the upper bound. One can check that the described trees of the theorem touches the above upper bound. Moreover, other cases with their respective values cannot reach this upper bound. The lower bound is clear since for any given ${}_{xy}T^k$, by the theorem 1, $D'({}_{xy}T^k) \geq 1$, as we had,

$$D'({}_{xy}T^k) = \sum_{\substack{(u,v)\in C_x\times C_y \\ d_T(u,v)>\frac{k}{2}}} (2d_T(u,v) - k).w_u.w_v.$$

On the other hand, if either of $w_u,\ w_v\ or\ (2d_T(u,v) - k)$ for some cases is greater than 1 then $D'({}_{xy}T^k) > 1$. This means the only trees which can touch the lower bound are the describe case of the theorem. This completes the proof. ■

**Remark 2:** The previous theorem gives the best upper bound of $D'({}_{xy}T^k)$ where $V(T) = n$ and $n$ is divisible by 8. Using the theorems' proof, it is possible establish a lower tight upper bound when $n \not\equiv mod\ 8$. Generally, if $D'({}_{xy}T^k)$ with $|V(T)| = n$ is maximum then:

$$k = \frac{n}{2} + i, \qquad i = 1 \text{ or } 2 \text{ or } 3 \text{ or } 4.$$

and

$$w_x = \left\lfloor \frac{n-k+2}{2} \right\rfloor, \qquad w_y = \left\lceil \frac{n-k+2}{2} \right\rceil \text{ and } \ w_v = 1,\ \ \forall v \in V(C)/\{x,y\}.$$

Note that as in theorem 2, $T_x$ and $T_y$ are arbitrary trees. ▲

Suppose we are given a ${}_{xy}T^k$ which means the inset edge is $x_1y_1$. If we remove $x_1y_1$ and substitute with $x_2y_2$ then we have a $(k-2)$-cycle in ${}_{x_2y_2}T$. Precisely, we can make just one change to the vectors ${}^{x}_{xy}\boldsymbol{W}$ and ${}^{y}_{xy}\boldsymbol{W}$ and calculate $D'({}_{x_2y_2}T)$. This lets us to prevent recalculating and thus we can save our time. In the next two theorems we present a variation of this fact.

**Theorem 3:** Suppose $T$ is a tree and the vectors ${}^{x}_{xy}\boldsymbol{W}$, ${}^{y}_{xy}\boldsymbol{W}$ and the matrix ${}_{xy}\boldsymbol{W} = [w_{ij}]$ are associated to ${}_{xy}T^k$, where $d_T(x,y) > 3$. Then $D'({}_{x_2y_2}T^{k-2}) \geq D'({}_{x_1y_1}T^k)$ if and only if:

$$\begin{cases} \displaystyle\sum_{\substack{4\le i+j\le k'+1\\ i,j>1}} \boldsymbol{w_{ij}} \ \ge\ \boldsymbol{w_{11}} & \boldsymbol{k} \textbf{ is even}, \\ \\ 2(\displaystyle\sum_{\substack{4\le i+j\le k'+1\\ i,j>1}} \boldsymbol{w_{ij}}) + \displaystyle\sum_{i+j=k'+2} \boldsymbol{w_{ij}} \ \ge\ \boldsymbol{2w_{11}} & \boldsymbol{k} \textbf{ is odd}. \end{cases}$$

**Proof:** Suppose we have ${}_{x_1y_1}T^k$ and $d_T(x_1, y_1) > 3$. If we remove $x_1y_1$ from ${}_{x_1y_1}T^k$ and add $x_2y_2$ to $T$ regarding the indexing of $C_{x_1}$ and $C_{y_1}$ in the ${}_{x_1y_1}T^k$ then we reach to ${}_{x_2y_2}T^{k-2}$. Therefore, we see that:

$${}^{x_2}_{x_2y_2}\boldsymbol{W} = {}^{x_1}_{x_1y_1}\boldsymbol{W}[2:] + {}^{x_1}_{x_1y_1}w_1.e_1,$$

and

$${}^{y_2}_{x_2y_2}\boldsymbol{W} = {}^{y_1}_{x_1y_1}\boldsymbol{W}[2:] + {}^{y_1}_{x_1y_1}w_1.e_1.$$

Moreover, we know that ${}_{x_2y_2}\boldsymbol{W} = {}^{y_2}_{x_2y_2}\boldsymbol{W} \times ({}^{x_2}_{x_2y_2}\boldsymbol{W})^{\boldsymbol{t}}$ . Therefore, using lemma 1, $D'({}_{x_2y_2}T^{k-2}) - D'({}_{x_1y_1}T^k) = \|\boldsymbol{F}_{k-2} \circledcirc {}_{x_2y_2}\boldsymbol{W}\| - \|\boldsymbol{F}_k \circledcirc {}_{x_1y_1}\boldsymbol{W}\|$ that is:

$$D'({}_{x_2y_2}T^{k-2}) - D'({}_{x_1y_1}T^k) = \begin{cases} \displaystyle\sum_{\substack{4\le i+j\le k'+1\\ i,j>1}} w_{ij} \ -\ w_{11} & k \text{ is even}, \\ \\ 2(\displaystyle\sum_{\substack{4\le i+j\le k'+1\\ i,j>1}} w_{ij}) + \displaystyle\sum_{i+j=k'+2} w_{ij} \ -\ 2w_{11} & k \text{ is odd}. \end{cases}$$

This completes the proof. ■

In the next theorem we compare $D'({}_{x_1y_1}T)$ and $D'({}_{x_2y_1}T)$ regarding the indexing of $C_{x_1}$ and $C_{y_1}$. That is, we remove the $x_1y_1$ from ${}_{x_1y_1}T$ and insert the edge $x_2y_1$ to the $T$.

**Theorem 4:** Suppose $T$ is a tree and the vectors ${}^{x}_{xy}\boldsymbol{W}$, ${}^{y}_{xy}\boldsymbol{W}$ and the matrix ${}_{xy}\boldsymbol{W} = [w_{ij}]$ are associated to ${}_{xy}T^k$, where $d_T(x, y) > 2$. Then $D'({}_{x_2y_1}T^{k-1}) \ge D'({}_{x_1y_1}T^k)$ if and only if:

$$\begin{cases} \sum_{\substack{i+j\leq k'+1 \\ i>1,\ j<k'}} \boldsymbol{w_{ij}} \geq \sum_{j<k'} \boldsymbol{w_{1j}} & \boldsymbol{k} \textbf{ is even}, \\ \sum_{\substack{i+j\leq k'+1 \\ i>1,\ j<k'}} \boldsymbol{w_{ij}} \geq \sum_{j\leq k'} \boldsymbol{w_{1j}} & \boldsymbol{k} \textbf{ is odd}. \end{cases}$$

**Proof:** Suppose $x$ and $y$ are two non-neighbor vertices of a tree $T$ with $d_T(x,y) > 2$. By lemma 1,

$$D'\left({}_{x_1y_1}T^k\right) = \left\| F_k \circledcirc \left({}^{x_1}_{x_1y_1}\boldsymbol{W}\right).\left({}^{y_1}_{x_1y_1}\boldsymbol{W}\right)^t \right\|, \tag{13}$$

and

$$D'\left({}_{x_2y_1}T^{k-1}\right) = \left\| F_{k-1} \circledcirc \left({}^{x_2}_{x_2y_1}\boldsymbol{W}\right).\left({}^{y_1}_{x_2y_1}\boldsymbol{W}\right)^t \right\|. \tag{14}$$

Note that changing the inset edge from $x_1y_1$ to $x_2y_1$ reduces the length of cycle of respective graphs by 1. And remember that the $\boldsymbol{F_k}$ matrices are different depends on whether $k$ is odd or even. This is very interesting to see that for both even and odd $k$:

$$\left\| F_{k-1} \circledcirc \left({}^{x_2}_{x_2y_1}\boldsymbol{W}\right).\left({}^{y_1}_{x_2y_1}\boldsymbol{W}\right)^t \right\| = \left\| (F_k + O_k) \circledcirc \left({}^{x_1}_{x_1y_1}\boldsymbol{W} - {}^{x_1}_{x_1y_1}w_1.e_1 + {}^{x_1}_{x_1y_1}w_1.e_2\right).\left({}^{y_1}_{x_1y_1}\boldsymbol{W}\right)^t \right\|.$$

The RHS of the above equation is equal to $D'\left({}_{x_2y_1}T^{k-1}\right)$. Expanding the LHS for odd and even cases separately, we can see that:

$$D'\left({}_{x_2y_1}T^{k-1}\right) - D'\left({}_{x_1y_1}T^k\right) = \begin{cases} \sum_{\substack{i+j\leq k'+1 \\ i>1,\ j<k'}} w_{ij} - \sum_{j<k'} w_{1j} & k \text{ is even}, \\ \sum_{\substack{i+j\leq k'+1 \\ i>1,\ j<k'}} w_{ij} - \sum_{j\leq k'} w_{1j} & k \text{ is odd}. \end{cases}$$

Therefore $D'({}_{x_2y_1}T^{k-1}) \geq D'({}_{x_1y_1}T^k)$ if and only if:

$$\begin{cases} \sum_{\substack{i+j\leq k'+1 \\ i>1,\ j<k'}} w_{ij} \geq \sum_{j<k'} w_{1j} & k \text{ is even}, \\ \\ \sum_{\substack{i+j\leq k'+1 \\ i>1,\ j<k'}} w_{ij} \geq \sum_{j\leq k'} w_{1j} & k \text{ is odd}. \end{cases}$$

This completes the proof. ■

In the next results we present an application of the theorems 3 and 4 to see that we can almost ignore the leaves of a tree $T$ in order to find $\max_{xy\in E(T^c)} D'({}_{xy}T)$.

**Corollary 2:** Suppose $T$ is a tree on $n > 6$ vertices and $T \ncong S_n$. Then there is a non-leaf vertex $v \in V(T)$ such that $D'({}_{xv}T) = \max_{xy\in E(T^c)} D'({}_{xy}T)$.

**Proof:** Suppose we are given a ${}_{xy}T^k$ such that $deg(x) = deg(y) = 1$. This means $w_x = w_y = 1$. If $k > 5$ then by theorem 3, $D'({}_{x_2y_2}T^{k-1}) \geq D'({}_{x_1y_1}T^k)$. If $k = 4$ then by theorem 4, $D'({}_{x_2y_1}T^{k-1}) \geq D'({}_{x_1y_1}T^k)$. If $k = 5$ and $|V(T)| > 6$ by theorem 4 either $D'({}_{x_2y_1}T^{k-1})$ or $D'({}_{x_1y_2}T^{k-1})$ or $D'({}_{x_My_1}T^{k-1})(D'({}_{x_My_1}T^{k-1}) = 1 + w_{x_2} + w_{x_M})$ is greater than $D'({}_{x_1y_1}T^k)$. If $k = 3$ the problem is clear by the Theorem 2. ■

**Theorem 5:** Suppose $v$ is a leaf of a tree $T$. Then for every $z \in V(T)$, $d_T(v,z) \in \mathbb{N}/\{2,3,4,6\}$, there is $y \in V(T)$ such that $y$ is not a leaf and $D'({}_{vz}T) \leq D'({}_{yz}T)$.

**Proof:** We break the proof to two claims which cover the theorem conditions.

**Claim (1):** If $v$ is an arbitrary leaf of a tree $T$, $z \neq v$, $d_T(v,z) = k-1$ and $k > 4$ is an even number then there is a non-leaf $u \in V(T)$ such that $D'({}_{vz}T) \leq D'({}_{uz}T)$.

**Proof of Claim (1):** By theorem 4's proof:

$$D'({}_{v_1z_1}T^k) - D'({}_{v_2z_1}T^{k-1}) = \sum_{\substack{i+j \le k'+1 \\ i>1,\ j<k'}} w_{ij} - \sum_{j<k'} w_{1j}\,, \tag{15}$$

Since $\deg(v) = 1$ so ${}^{v}_{vz}w_1 = 1$. Therefore,

$$\sum_{j<k'} {}^{z}_{vz}w_j = \sum_{j<k'} w_{1j}\,, \tag{16}$$

and for $k' \geq 3$,

$$\sum_{j<k'} {}^{v}_{vz}w_j \leq \sum_{\substack{i+j \le k'+1 \\ i>1,\ j<k'}} w_{ij,} \tag{17}$$

Thus using (15-17),

$$ED({}_{v_1z_1}T^{k-1}) \leq ED({}_{v_2z_1}T^{k-1}).$$

**Claim (2):** If $v$ is an arbitrary leaf of a tree $T$, $z \neq v$, $d_T(v,z) = k-1$ and $k > 7$ is an odd number then there is a non-leaf $u \in V(T)$ such that $D'({}_{vz}T) \leq D'({}_{uz}T)$.

**Proof of Claim (2):** The proof for the odd $k$'s is tricky and needs a precise comparison. First, we know that by the theorem 4's proof:

$$D'({}_{v_1z_1}T^k) - D'({}_{v_2z_1}T^{k-1}) = \sum_{\substack{i+j \le k'+1 \\ i>1,\ j<k'}} w_{ij} - \sum_{j \le k'} w_{1j,} \tag{18}$$

And by the theorem 3's proof:

$$D'({}_{v_1z_1}T^k) - D'({}_{v_2z_2}T^{k-2}) = 2\Big(\sum_{\substack{4 \le i+j \le k'+1 \\ i,j>1}} w_{ij}\Big) + \sum_{i+j=k'+2} w_{ij} - 2w_{ij}, \tag{19}$$

According to (18) and (19) respectively if $D'({}_{v_1z_1}T^k) \geq D'({}_{v_2z_1}T^{k-1})$, $D'({}_{v_1z_1}T^k) \geq D'({}_{v_2z_2}T^{k-2})$, ${}^{v}_{vz}\boldsymbol{W} = [1,1,\ldots,1]^t$ and ${}^{z}_{vz}\boldsymbol{W} = [a_i]$ then:

$$a_{k'} \geq \sum_{i=1}^{k'-1} (k'-i)a_i\,, \tag{20}$$

and

$$2a_1 \geq \sum_{i=2}^{k'} (2k' - 2i + 1) a_i. \quad (21)$$

It is straightforward to see that if $k' \geq 4$ then either of (20) or (21) does not hold. Meaning that if (20) holds then (21) does not happen and if (21) happens then (20) does not hold. Moreover, it is easy to see if ${}^{v}_{vz}\boldsymbol{W} \neq [1,1,\dots,1]^t$ then the RHS of (20) and (21) will be greater and this does not harm our argument. Consequently using (18) and (19), either $D'({}_{v_1z_1}T^k) \geq D'({}_{v_2z_1}T^{k-1})$ or $D'({}_{v_1z_1}T^k) \geq D'({}_{v_2z_2}T^{k-2})$ which completes the proof of claim of 2 and therefore the theorem's proof. ■

**Remark 3:** It is not hard to take care of the exceptions of theorem 5 by specifying the number of vertices and diameter of the input trees. Our ultimate goal is to apply theorems 3 and 4 through an algorithm to find $\{D'({}_{xy}T)\}_{xy\in V(T^c)}$ and $\max_{xy\in E(T^c)} D'({}_{xy}T)$. However, by theorem 5 and corollary 2, using the facts that for a tree $T$ on $n$ vertices $|E(T^c)| \approx \frac{n^2}{2}$ and the following result by Renyi:

**Proposition 1:** The expected number of leaves in a random labeled tree on $n$ vertices is $n/e$ where $e$ is the base of the natural number. ■

we can limit our search space from $\binom{n}{2} - |E(T)|$ inset edges to almost $\binom{n-\frac{n}{e}}{2}$ inset edges, where we are looking for the $\min_{xy\in E(T^c)} AD({}_{xy}T)$. This means we can prone $1 - \frac{\binom{n-\frac{n}{e}}{2}}{\binom{n}{2}} \approx 1 - \left(\frac{e-1}{e}\right)^2 \approx \%60$ of edges in average, in sake of $\min_{xy\in E(T^c)} AD({}_{xy}T)$. ▲

The next theorem shows how to apply theorems 3 and 4 through an algorithm.

**Theorem 6:** Suppose $T$ is a tree, $xy \in V(T^c)$ and $d_T(x,y) = k-1$. If we are given the vectors ${}^{x}_{xy}\boldsymbol{W}$ and ${}^{y}_{xy}\boldsymbol{W}$ then we can obtain

$$\left\{D'({}_{x_iy_i}T^k)\right\}_{1\leq i\leq k'-1} \cup \left\{D'({}_{x_{i+1}y_i}T^k)\right\}_{1\leq i\leq k'-1} \cup \left\{D'({}_{x_iy_{i+1}}T^k)\right\}_{1\leq i\leq k'-1},$$

with $O(k^2)$ operations.

**Proof:** Suppose for a tree $T$, $xy \in E(T^c)$ and the related vectors of ${}_{xy}T^k$ , ${}^{x}_{xy}\boldsymbol{W}$ and ${}^{y}_{xy}\boldsymbol{W}$**,** are given. Using Lemma 1:

$$D\left({}_{xy}T^k\right) = \left\|\boldsymbol{F}_k \circledcirc ({}^{x}_{xy}\boldsymbol{W} \times {}^{y}_{xy}\boldsymbol{W})\right\|,$$

which clearly costs us $O(k^2)$ operations to calculate $D\left({}_{xy}T^k\right)$, ${}_{xy}\boldsymbol{W}$ and $\left\|\boldsymbol{O}_{\boldsymbol{k}} \circledcirc {}_{xy}\boldsymbol{W}\right\|$. Now we prove that if we have $D'\left({}_{x_i y_i}T^k\right)$ and ${}_{xy}\boldsymbol{W}$ then we can obtain $D'\left({}_{x_{i-1} y_{i-1}}T^k\right)$ by at most $O(k)$ operations. Without loss of generality suppose we are given $D'\left({}_{x_1 y_1}T^k\right)$ then by theorem 3 if $k$ is even,

$$D'\left({}_{x_2 y_2}T^{k-2}\right) - D'\left({}_{x_1 y_1}T^k\right) = \sum_{\substack{4 \le i+j \le k'+1 \\ i,j>1}} w_{ij} - w_{11}, \tag{22}$$

Since we already have $\left\|\boldsymbol{O}_k \circledcirc {}_{xy}\boldsymbol{W}\right\|$, this is enough to subtract the first row and column of ${}_{xy}\boldsymbol{W}$ from it to obtain $D'\left({}_{x_2 y_2}T^{k-2}\right)$. This means,

$$\sum_{\substack{4 \le i+j \le k'+1 \\ i,j>1}} w_{ij} - w_{11} = \left\|\boldsymbol{O}_k \circledcirc {}_{xy}\boldsymbol{W}\right\| - {}^{x}_{xy}w_1\left(\sum_{1 \le i \le k'} {}^{y}_{xy}w_i\right) - {}^{y}_{xy}w_1\left(\sum_{1 \le i \le k'} {}^{x}_{xy}w_i\right) \tag{23}$$

Using (22, 23) we can obtain $D'\left({}_{x_2 y_2}T^{k-2}\right)$ from the given data, with $O(k)$ operations as well as calculating $\left\|\boldsymbol{O}_{k-2} \circledcirc {}_{x_2 y_2}\boldsymbol{W}\right\|$ and ${}_{x_2 y_2}\boldsymbol{W}$. Correspondingly, we can obtain $D'\left({}_{x_3 y_3}T^{k-4}\right)$ with $O(k)$ operations. Therefore, we obtained $\left\{D'\left({}_{x_i y_i}T^k\right)\right\}_{1 \le i \le k'-1}$ in $O(k^2)$. A similar argument applies for odd $k$'s.

Similarly, using Theorems 4 we can reach to $\left\{D'\left({}_{x_{i+1} y_i}T^k\right)\right\}_{1 \le i \le k'-1}$ and $\left\{D'\left({}_{x_i y_{i+1}}T^k\right)\right\}_{1 \le i \le k'-1}$ in $O(k^2)$. This completes the proof. ■

**Remark 4:** The average diameter of a tree $T$ on $n$ vertices is $log\,(n)$ [8]. On the other hand, by theorem 6 we can reach to $\{D'(_{x_iy_i}T^k)\}_{1\leq i\leq k'-1}$ in $O(k^2)$. Thus, in average, we could obtain each $D'(_{x_iy_i}T^k)$, $1 \leq i \leq k' - 1$, in $O(k)$, subject to the given condition. Since $k < diam(T) + 1$ and the number of inset edges is $O(n^2)$ we have the following question:

**Question 1:** Suppose $T$ is a tree on $n$ vertices. Is it possible to obtain $\{AD(_{xy}T)\}_{xy\in E(T^c)}$ with the average time complexity of $O(n^2 \log(n))$?